\documentclass[aps,pre,floats,twocolumn,showpacs,superscriptaddress]{revtex4}

\usepackage{graphicx,epsfig}
\usepackage{times}
\usepackage{graphics,dcolumn,bm,fleqn,float}
\usepackage{amssymb,amsmath,multirow,rotate,color}
\begin{document}

\title{Social Network Reciprocity as a Phase Transition in Evolutionary Cooperation.}

\author{L. M. Flor\'{\i}a}

\email{mario.floria@gmail.com}

\affiliation{Institute for Biocomputation and Physics of Complex
Systems (BIFI), University of Zaragoza, Zaragoza 50009, Spain}

\affiliation{Departamento de F\'{\i}sica de la Materia Condensada,
University of Zaragoza, Zaragoza E-50009, Spain}

\author{C. Gracia-L\'azaro}

\affiliation{Departamento de F\'{\i}sica de la Materia Condensada,
University of Zaragoza, Zaragoza E-50009, Spain}

\author{J. G\'{o}mez-Garde\~{n}es}

\affiliation{Institute for Biocomputation and Physics of Complex
Systems (BIFI), University of Zaragoza, Zaragoza 50009, Spain}

\affiliation{Department of Computer Science and Mathematics, University Rovira i Virgili, 
43007 Tarragona, Spain}

\author{Y. Moreno}

\email{yamir.moreno@gmail.com}

\affiliation{Institute for Biocomputation and Physics of Complex
Systems (BIFI), University of Zaragoza, Zaragoza 50009, Spain}

\affiliation{Departamento de F\'{\i}sica Te\'orica. University of
Zaragoza, Zaragoza E-50009, Spain}

\date{\today}

\begin{abstract}

In Evolutionary Dynamics the understanding of cooperative
phenomena in natural and social systems has been the subject of
intense research during decades. We focus attention here on the
so-called "Lattice Reciprocity" mechanisms that enhance
evolutionary survival of the cooperative phenotype in the
Prisoner's Dilemma game when the population of darwinian
replicators interact through a fixed network of social contacts.
Exact results on a "Dipole Model" are presented, along with a
mean-field analysis as well as results from extensive numerical
Monte Carlo simulations. The theoretical framework used is that of
standard Statistical Mechanics of macroscopic systems, but with no
energy considerations. We illustrate the power of this perspective
on social modeling, by consistently interpreting the onset of
lattice reciprocity as a thermodynamical phase transition that,
moreover, cannot be captured by a purely mean-field approach.

\end{abstract}


\maketitle

\section{Introduction}
\label{I}

Is the term "social temperature" just a rhetoric figure
(suggestive metaphor), or on the contrary, could it be given a
precise meaning? By working out in detail the Evolutionary
Dynamics of the most studied social dilemma (the Prisoner's
Dilemma) on a simple kind of artificial social networks we will
show here that the formal framework of Equilibrium Statistical
Mechanics is, to a large extent, applicable to the rigorous
description of the asymptotic behavior of strategic evolution,
thus providing the key for a formal quantitative meaning of the
term social "temperature" in these contexts.

Evolutionary game theory, in contrast with classical game theory
that focusses on the decision making process of (rational) agents,
is concerned with entire populations of agents programmed to use
some strategy in their interactions with other agents. The agents
are replicators, {\em i.e.} entities which have the means of
making copies of themselves (by inheritance, learning, infection,
imitation, etc...), whose reproductive success depends on the
payoff obtained during interaction. As the payoff depends on the
current composition of strategies among the interacting agents,
this yields a feedback loop that drives the evolution of the
strategic state of the population
\cite{hofbauer,gintis,hofbauer03,nowak06}.

This Darwinian feedback (frequency-dependent fitness) dynamics
depends strongly not only on the particular game, and on the
specifics of the way strategies spread, but also on the (social)
structure of connections describing the interactions. Under the
assumption of a well-mixed population ({\em social panmixia}
assumption), the temporal evolution of the proportion of
strategies among the population is governed by a differential
equation named {\em replicator equation} (see below). Well-known
celebrated folk's theorems (see, e.g. \cite{hofbauer03}) establish
a connection between the asymptotic behavior of this equation and
the powerful concepts of classical game theory based on the notion
of best reply (Nash). However, if the social panmixia assumption
is abandoned, and individuals only interact with their neighbors
in a social network, the asymptotic of evolutionary dynamics
generically differ in a substantial way from this "well-mixed
population" description. The social structure of strategic
interactions turns out to be of importance regarding the
evolutionary outcome of the strategic competition.

We will consider here the Prisoner's Dilemma (PD), a
two-players-two-strategies game, where each player chooses one of
the two available strategies, cooperation or defection: A
cooperator receives $R$ when playing with a cooperator, and $S$
when playing with a defector, while a defector earns $P$ when
playing with a defector, and $T$ (temptation) against a
cooperator. When $T>R>P>S$, the game is a PD (while if $T>R>S>P$
it is called Snowdrift game, also "Chicken" or "Hawks and Doves").
Given the payoff's ordering, whatever the value of the prior
assign of probability to the co-player's strategy is, the expected
payoff is higher for defection, and that is what a rational agent
should choose. In the PD game only the defective strategy is a
strict best response to itself and to cooperation, thus it is an
easy example of game with an unbeatable \cite{nowak06} strategy.
Still, though there is no difficulty in the making of the
strategic decision from Nash analysis, two cooperators are better
off than two defectors, hence the Social Dilemma.

In graph-structured populations, a large body of research
\cite{nowak,ohtsuki,abramson,maxi,duran,hungarians,szabo,tpn07} (and
references therein) on evolutionary dynamics of the PD game has
convincingly show the so-called {\em lattice reciprocity} effects:
The cooperative phenotype can take advantage of the topology of
the social net, so that clusters of cooperators are often
resilient to invasion by the (continuum-unbeatable) defective
phenotype. This enhancement of asymptotic macroscopic levels of
cooperation due to the structure and topology of strategic
interactions includes, but it is far more general than, the
so-called space reciprocity mechanisms, where social nets are
discretizations (solid state lattices) of the euclidian space, and
diffusion approximations are often useful \cite{Hauert2002}. In
this regard, one should stress the accumulated evidence that ({\em
i}) many interesting social nets \cite{newmanrev,yamirrep,bara}
are far away from being regular lattices, and ({\em ii}) freedom
of connectivity scales (scale-free complex networks) enhances
\cite{pacheco2,us,njp,jtb} the lattice reciprocity mechanisms up
to unexpectedly high values of the temptation parameter $T$ of the
dilemma, where cooperation is very expensive (but affordable in an
evolutionary sense).

In this paper we investigate in detail the lattice reciprocity
mechanisms in an artificial network (Dipole Model) that models the
competition for influence on a population of social PD-imitators
of two antagonist Big Brothers (nodes connected to the whole
population, but with no direct connection between them). The paper
is organized as follows. The setting of evolutionary scenario [(a)
game, (b) updating rule, and (c) network of social contacts] along
with basic concepts and definitions, are given in section \ref{ED
in graphs}, where the Dipole Model is introduced. This is a closed
system with a self-sustained evolutionary activity (non-trivial
dynamics) of social cooperation, as we prove in section
\ref{dipole_net}, where also the applicability of standard
Equilibrium Statistical Mechanics to the Dipole Model is assessed.

Explicit exact solutions of the evolutionary equilibrium
probability measure of microstates are obtained in section
\ref{asymptotic}, for some special topologies of the fluctuating
replicators subpopulation ${\cal F}$. There, we analyze first the
two trivial limits for the graph structure of the target
population ${\cal F}$, namely complete graph and totally
disconnected graph. One easily obtains an exact macroscopic
(infinite size limit, or thermodynamical limit) description for
both cases by means of explicit differential equations for the
macroscopic cooperation in subsections \ref{panmixia} and
\ref{ideal gas}. A simple thermodynamical interpretation of the
macroscopic behavior, is provided by the theorem of section
\ref{dipole_net}, as we briefly outline. For a simple "random
regular graph" topology of ${\cal F}$, an explicit differential
equation and simple thermodynamical predictions are obtained
within a mean-field approximation in subsection \ref{mean-field}.
When compared to Monte Carlo numerical results, fundamental
discrepancies are evidenced: while mean-field prediction does not
show any critical behavior, our numerical results show beyond any
doubt the existence of a thermodynamical phase transition at a
critical value of the temptation $T^*$. This critical value
separates apart two distinct macroscopic phases of the fluctuating
population, and signals the onset of macroscopic effects of
lattice reciprocity. These effects are seen to operate as positive
feedback upon local fluctuations of the strategic neighborhoods,
and thus they cannot be captured by "purely mean-field"
macroscopic approaches.

The concluding section \ref{IV} tries to call interdisciplinary
attention on the wide and utmost interesting prospectives for
Statistical Physics "concepts and methods" in current studies on
Evolutionary Dynamics and Social Systems modelling in general.

\section{Natural strategic selection on graphs.}
\label{ED in graphs}

We specify here the evolutionary game dynamics scenario, meaning
the game parametrization, the microscopic strategic dynamics
(replication mechanism or strategic updating rule), and the social
structure of contacts that we will consider along the paper.

We normalize the PD payoffs to the reward for cooperating, $R=1$,
and fix the null payoff at punishment $P=0$. Note that provided
the (differential or relative) selective advantage among two
individuals depends on their payoff's difference (see below), one
can arbitrarily fix the zero payoff level. Then only two
parameters $T=b>1$ and $R=\epsilon<0$ are tuned. Note that the
range $\epsilon>0$ defines a game named Hawks and Doves (also
Chicken and Snowdrift) where punishment and sucker's payoff have
the reverse order. We will occasionally comment on this range of
parameters.

Moreover, we do not restrict our computations to $2R>T+S$. This
restriction means that the total payoff for the two players is
higher if both cooperate ($2R$) than if one cooperates and the
other defects ($T+S$), and is usually incorporated in iterated
games studies of the PD to prevent agents taking turns at
defection and then sharing the payoffs. For the specifics of the
replicator dynamics (memory-less, markovian) in the next
paragraph, one should not expect that this restriction
qualitatively matters.

Regarding the replication mechanism, we implement the finite
population (size $N \gg 1$) analogue of replicator dynamics
\cite{weibull,pacheco2}. At each time step $t$, which represents
one generation of the discrete evolutionary time, each agent $i$
plays once with each one of the agents in its neighborhood and
accumulates the obtained payoffs, $P_i$. Then, the individuals,
$i$, update synchronously their strategies by picking up at random
a neighbor, $j$, and comparing their respective payoffs $P_i$ and
$P_j$. If $P_i > P_j$, nothing happens and $i$ keeps the same
strategy for the next generation. On the contrary, if $P_j > P_i$,
with probability $\Pi_{i\rightarrow j}=\beta(P_j-P_i)$, $i$ adopts
the strategy of its neighbor $j$ for the next round robin with its
neighbors, before which all payoffs are reset to zero. Here
$\beta$ is a number small enough to make $\Pi_{i\rightarrow j}$ an
acceptable probability; its physical meaning is related to the
characteristic inverse time scale: the larger it is, the faster
evolution takes place.

From a theoretical point of view, this specific choice of the
dynamics has the virtue of leading directly (see, {\em e.g.}
\cite{gintis}), under the hypothesis of a well-mixed population
and very large population size, to the celebrated replicator
equation for the frequencies $p_{\alpha}$ of strategies $\alpha$(=
C or D) in the population:

\begin{equation}
\dot{p}_{\alpha} = p_{\alpha}(f_{\alpha}-\bar{f})
\label{replicator}
\end{equation}
where $f_{\alpha}$ is the payoff of an $\alpha$-strategist and
$\bar{f}$ is the average payoff for the whole population. Note
that time unit in equation (\ref{replicator}) is scaled to
$\beta^{-1}$.

For the payoffs of the Prisoner's Dilemma the asymptotic frequency
of cooperators, from the replicator equation, is driven to
extinction, $p_c = 0$, while for the Hawks and Doves game, its
asymptotic value is $\epsilon/(b-1+\epsilon)$ . As stated in the
introductory section, we will be concerned here mainly with
populations that are not well-mixed, where predictions based on
this nonlinear differential equation are often of little use.

Regarding the structure of connections between interacting agents,
we will consider here that it is given by a fixed graph ({\em
i.e.} connections between players do not change by rewiring) where
agents are represented by nodes, and a link between nodes
indicates that they interact (play). If $k_i$ is the number of
neighbors of agent $i$ (connectivity or degree), and $\Delta$ is
the maximal possible one-shot-payoff difference ($\Delta =
\text{max}\{b,b-\epsilon\}$), we will assume $\beta =
(\text{max}\{k_i,k_j\} \Delta)^{-1}$ for the specification of the
probability $\Pi_{i\rightarrow j}$ of invasion of node $i$ by the
strategy of neighbor $j$. This simple choice, introduced in
\cite{pacheco2}, assures that $\Pi_{i\rightarrow j} < 1$; in
heterogeneous networks it has also the effect of slowing down the
invasion processes from or to highly connected nodes, with respect
to the rate of invasion processes between poorly connected nodes,
a feature not without consequences \cite{plosone}.

We now introduce some notation, which is familiar to statistical
physicists: The configuration ({\em strategic microstate} $l$) of
a population of $N$ agents at time $t$ is specified by the
sequence $l=\{s_i(t)\}$ ($i=1, ..., N$), where $s_i(t) = 1$ (or
$0$) denotes that node $i$ is at this time a cooperator (resp.
defector). The set of all possible $2^N$ configurations is called
the phase space. Stationary probability densities of microstates
${\cal P}(l)$ ($l=1,...2^{N}$) are then representatives of {\em
strategic macro-states}. The average cooperation $c_l$ of
microstate $l$ is defined as
\begin{equation}
c_l = \frac{1}{N} \sum_i^N s_i \label{cooperation}
\end{equation}

We denote by $\Pi_{l'l}$ the probability that the strategic
microstate of the population at time $t+1$ is $l'$, provided that
it is $l$ at time $t$. Note that $\sum_{l'} \Pi_{l'l}=1$. A
microstate $\hat{l}$ is a {\em frozen} equilibrium configuration
if the probability that it changes in one time step is null, and
then $\Pi_{\hat{l}\hat{l}}=1$ and $\Pi_{l' \hat{l}}=0$ if $l' \neq
\hat{l}$. We will assume generic real values (irrational) of the
payoff parameters, so that if a configuration contains a C-D link
it cannot be a frozen configuration. The only possible frozen
equilibrium configurations are {\em all}-C and {\em all}-D.
However, for a very wide class of graphs, and a wide range of
model parameters they are not the only possible stationary
probability measures.

We now illustrate by means of easy examples the evolution of PD on
graphs. Our first and simplest example is a star-shaped graph
consisting of a central node connected to $N-1$ peripheral nodes.
It is straightforward to check that any initial condition with
cooperators at the central node and (at least) at
Int$[(b-\epsilon(N-1))/(1-\epsilon)]+ 1$ peripheral nodes has a
positive probability of evolving in one time step to a
configuration with a higher number of cooperators, and a null
probability of evolving towards less cooperators. Thus, all those
configurations evolve asymptotically to the {\em all}-C
equilibrium. The rest of configurations evolve towards the {\em
all}-D equilibrium. Therefore, if $N>(b-\epsilon+2)$ both
equilibria are attractors (absorbing states), in the sense that
some configurations different from themselves evolve to them; the
phase space is partitioned into two basins of attraction. If $N <
b -\epsilon +2$, only the {\em all}-D frozen equilibrium is
attractor. The stationary probability densities ${\cal P}^*(l)$ of
the star are pure point measures (two- or one- Dirac delta peaks)
in the thermodynamical limit $N \rightarrow \infty$.

Now take a star and add some arbitrary number of links between its
peripheral nodes. We call this network a crown, whose head is the
central node. If the head is occupied at $t_0$ by a defector, it
will remain so forever, because the payoff of a peripheral
cooperator is strictly lower than head's payoff. Sooner or later
the head (center) of the crown will be imitated by the whole
crown, and the evolution will stop when everybody be defecting.
But, what happens to a cooperator on the head? The answer is
dependent on both, the net topology of the crown periphery and the
cooperators disposition there: To ensure fixation of cooperation
at the head node, it suffices that a subset $C$ of peripheral
nodes occupied by cooperators, and with no direct links to the
rest of the periphery, have a size $n_C > b
k_{max}-\epsilon(N-n_C-1)$, where $k_{max}$ is the maximal degree
in the rest of the periphery. Under this proviso all-$C$ is the
unique absorbing microstate of all corresponding initial
conditions.

\begin{figure}
\begin{center}
\epsfig{file=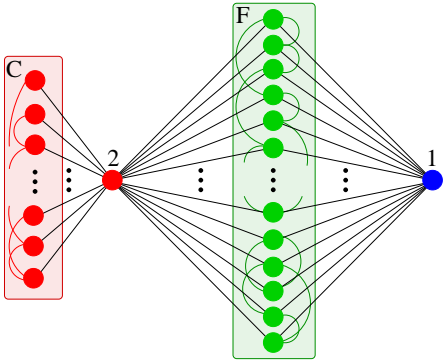,width=2.3in,angle=-0,clip=1}
\end{center}
  \caption{Structure of the Dipole Network. Two nodes ($1$ and $2$) are connected to all nodes in $\cal F$, whose elements can be arbitrarily linked to each other. Moreover, node $2$ is also linked to a set $\cal C$ (with arbitrary internal connections as well). See the text for further details.}
\label{bipolnet}
\end{figure}

Finally consider the graph schematized in Fig.\ \ref{bipolnet},
composed of the following: \begin{itemize} \item[({\em a})] A
component $\cal F$ of $n_F$ nodes with arbitrary connections among
them. \item[({\em b})] A node, say node 1, that is connected to
all the nodes in $\cal F$ and has no other links. \item[({\em c})]
A component $\cal C$ of $n_C$ nodes with arbitrary connections
among them. \item[({\em d})] A node, say node 2, that is connected
to all the nodes in $\cal F$ and $\cal C$, but not to node 1.
\end{itemize}

This is what we will call a Dipole Model Network. It is a
two-headed (nodes 1 and 2) crown (with periphery $\cal F$) plus a
tail $\cal C$ hanging on head 2. To strength the special status of
the head nodes, let us nickname them as "Big Brothers". They
certainly enjoy a sort of omnipresence that fits well with the
character of Orwell's famous social sci-fiction novel $1984$. In
the following section we prove that for this simple network there
exists a non-trivial stationary probability density of microstates
${\cal P}^*(l)$ for the strategic evolution of the PD game.

\section{The Dipole Model.}

\label{dipole_net}

The analysis of Evolutionary Dynamics of the PD on the Dipole
network shows that there is a non-trivial invariant measure in
phase space. Let us consider the set $\cal I$ of initial
conditions defined by: ({\em i}) Big Brother 1 is a defector,
({\em ii}) Big Brother 2 is a cooperator, and ({\em iii}) all
nodes in component $\cal C$ are cooperators. Note that this set
contains $2^{n_F}$ different configurations. We now prove that,
provided some sufficient conditions, this is a minimally invariant
set of the evolutionary dynamics.

First, one realizes that Big Brother 1 cannot be invaded by the
cooperative strategy: The payoff of a cooperator node $i$ in $\cal
F$ is $P_i^c=k_i^c+1+\epsilon(k_i-k_i^c+1)$, where $k_i$ is the
number of its neighbors in $\cal F$ and $k_i^c\leq k_i$ is the
number of those that are cooperators. The payoff of Big Brother 1
(BB1) is then $P_1\geq(k_i^c+1)b$. For the PD game, where
$\epsilon \leq 0$, the inequality $P_1>P_i^c$ always holds, so
that BB 1 will always be a defector. (Note also that for the Hawks
and Doves game, a sufficient condition for $P_1>P_i^c$ is
$b>1+\epsilon(k_F+1)$, where $k_F$ ($<n_F$) is the maximal degree
in component $\cal F$, {\em i.e.} the maximal number of links that
a node in $\cal F$ shares within $\cal F$.) We thus conclude that
defection is fixed at BB 1.

Second, thanks to its interaction with set ${\cal{C}}$, Big
Brother 2 resists invasion, provided its size $n_C$ is above a
threshold: The payoff of a defector node $i$ in $\cal F$ is
$P_i^d=(k_i^c+1)b$, where $k_i^c$ is the number of its cooperator
neighbors in $\cal F$, while the payoff of Big Brother 2 (BB2) is
$P_2=n_C + n_F\epsilon + n_F^c(1-\epsilon)$, where $n_F^c\leq n_F$
is the number of cooperators in $\cal F$. Thus, a sufficient
condition for $P_2>P_i^d$ is $n_C>
\mathrm{Int}(b(k_F+1)-n_F\epsilon)$. With this proviso, BB2 will
always be a cooperator, which in turn implies that all the nodes
in the component $\cal C$ will remain always cooperators.

The previous argument proves that provided the sufficient
conditions $n_C> \mathrm{Int}(b(k_F+1)-\epsilon n_F)$ and $b > 1 +
\epsilon (k_F+1)$ hold, the subset $\cal I$ of phase space defined
by ({\em i}), ({\em ii}), and ({\em iii}) is an invariant set. As
this set does not contain equilibria, no stochastic trajectory
evolves from it to a frozen equilibrium configuration.

Finally, one realizes that ${\cal{I}}$ is indeed minimal, because
at any time, a defector in $\cal F$ has a positive probability to
be invaded by the cooperation strategy (from BB2), and a
cooperator in $\cal F$ has a positive probability of being invaded
by the defection strategy (from BB1). Therefore, any strategic
configuration of the set $\cal I$ is reachable in one time step
from any other, {\em i.e.} for all pairs ($l$, $l'$) of
microstates in $\cal I$, the transition probability $\Pi_{l'l}>0$.
Consequently, $\cal I$ does not contain proper invariant subsets:
it is minimally invariant. Moreover, following Perron-Frobenius
theorem, there exists a unique equilibrium macro-state ${\cal
P}^*(l)$. This provides a rigorous framework for the
interpretation of results from numerical Monte Carlo simulation
studies in Evolutionary Dynamics on Dipole models, provided the
sufficient conditions above.

While nodes in $\cal C$ and Big Brother 2 are permanent
cooperators, and Big Brother 1 is a permanent defector, nodes in
$\cal F$ are forced to fluctuate. This partition of the network
into sets of nodes where each particular strategy is fixed
forever, and a set of fluctuating nodes, turns out to be a generic
feature of the discrete replicator dynamics (neighbor imitation
proportional to payoffs difference) on many network settings
\cite{us,njp}. The simplicity of the Dipole Network model allows
on it an easy formal proof of existence of this partition, so
providing an illustration of both, its origins and generic
character. It also shows the formal applicability of Equilibrium
Statistical Physics formalism to characterize the asymptotic
behavior of Evolutionary Dynamics on these graphs. This will be
made in the next section for specific choices of structural traits
for the subgraph ${\cal F}$.

The name dipole for this structure of connections is suggested by
the strategic polar (${\cal C}- {\cal F}- {\cal D}$) aspect of the
whole graph. Note also that the number of ${\cal C}-{\cal F}$ and
${\cal F}-{\cal D}$ connections scales linearly with the size
$n_F$ of the fluctuating interior, that is to say that the poles
(C and D) act as an externally imposed (AC) field on $\cal F$,
whose strength is proportional to the internal levels of
cooperation. As the cooperation (and then the fitness) levels are
self-sustained (as proved by the previous theorem), this is a
closed macroscopic system with a non-trivial self-sustained social
activity of cooperation at evolutionary equilibrium.

The interest of the Dipole Model is by no means restricted to a
mere academic illustration: First of all, we can make a technical
use of it in macroscopic stability analysis studies of
PD-evolution on highly heterogeneous complex networks. Indeed, the
fluctuations inside the subset $\cal F$ are the effect of the
competition for invasion among two non-neighboring hubs (hugely
connected nodes), where opposite pure strategies have reached
fixation, in their common neighborhood. This is a local strategic
configuration that mimics those that are often observed in
stochastic simulations of evolutionary dynamics in highly
heterogeneous (scale-free) networks \cite{us,njp}. Simple
multipolar network models can easily be constructed (e.g. by
establishing direct links from $\cal C$ to $\cal F$ in a way that
simple sufficient conditions guarantee that the theorem still
holds), that are indeed indistinguishable from typical strategic
patterns found in the numerical simulations on scale-free
networks. This makes the Dipole net a very useful technical device
to analyze the stability mechanisms of the cooperator clusters
\cite{us,njp} in scale-free structured populations, as well as the
kind of temporal fluctuations of cooperation that one should
expect in the fluctuating set of nodes.

Regarding potentialities for Econo-Socio-Physics applications of
the Dipole model, it could be viewed as a sort of schematic (then
simplistic, cartoon-like) model for the competition for influence
of two powerful superstructural institutions ({\em e.g.} like
"mass media", political parties, or lobbies) on a target
population, in strongly polarized strategic contexts. The analysis
rigorously provides sufficient conditions for the parameter values
where fixation of strategic traits is proved impossible, so that
temporal fluctuations dominate forever the target population of
social imitators $\cal F$. The influence on each individual of the
two competing institutions is simulated here through the
omnipresent ("Big Brother" nodes 1 and 2) neighbors, whose own
high appeal for imitation (the strength of Big Brother's
influence) is in turn conditioned by the strategic composition of
the target population. Here the interest could well be the study
of the influence that metric and topological network
characteristics of the social structure have on the strategic
macro-state, and thus on the quantitative values of {\em social
indicators}. We address some aspects of this issue in the next
section.

At a more general level, the design of experiments in Social
Sciences as well as theoretical studies of Artificial Societies
could greatly benefit from having at hand simple but non-trivial
"exactly soluble statistical-mechanical models" that may provide
safe guides to develop further intuitions on social phenomena that
demands more comprehension.

\section{The role of social structure in Big Brothers competition.}

\label{asymptotic}

In this section we present some analytical and numerical results
on the evolutionary dynamics of games in the Dipole Model for
different choices of topologies of the fluctuating set $\cal F$.
The sufficient conditions stated in the previous section are
assumed hereafter. We are interested in the situation where
$n_F\gg1$, {\em i.e.} large size of the fluctuating population.
First we will analyze in \ref{panmixia} the straightforward
limiting case when the macroscopic set $\cal F$ is a fully
connected set. This is the well-mixed population limit, where
replicator equation is an exact description. Next in subsection
\ref{ideal gas} we will explicitly solve the opposite trivial case
of disconnected $\cal F$ set ($k_F=0$), which turns out to reduce
to the standard textbook ideal two-states model of Statistical
Physics. After that, in subsection \ref{mean-field}, the "random
regular" network structure for $\cal F$ is seen to be amenable to
a plausible mean-field approach, but insufficient to explain the
phenomenology shown by Monte Carlo numerical results. These show
beyond any doubt a critical behavior, a transition point
separating two qualitatively different types of social
macro-states. This transition is sensibly interpreted as the onset
of lattice reciprocity. In other words, {\em lattice reciprocity
is a true critical social phenomenon}.

\subsection{Well-mixed population limit.}
\label{panmixia}

 We now show that the well-mixed population assumption in the
macroscopic set $\cal F$ leads in the thermodynamic limit to the
standard replicator equation: Assume that each node in $\cal F$ is
connected to all nodes in $\cal F$, and to BB1 and BB2. The degree
of each of them is thus $n_F+1$, and the parameter $k_F=n_F-1$.
The sufficient conditions for fixation of defection at node 1 and
of cooperation at node 2 are respectively, $b>1+\epsilon n_F$
(that always hold for the PD), and
$n_C>\mathrm{Int}(n_F(b-\epsilon))$. The payoffs of polar nodes 1
and 2 are given by

\begin{equation}
P_1= bcn_F \;\;\;,\;\;\;\;\;\;\;\;\;P_2=n_C + cn_F +\epsilon(1-c)
n_F\;\;, \label{payoff12}
\end{equation}
while the payoffs of a cooperator node and a defector node in
$\cal F$ are
\begin{equation}
P_c= cn_F + \epsilon(n_F -cn_F +1)
\;\;\;\;,;\;\;\;\;\;\;P_d=(cn_F+1)b\;\;. \label{payoffCDbis}
\end{equation}

To compute the invasion probabilities, $Q_{DC}$ and $Q_{CD}$, one
first easily realizes that $P_c<P_d$, provided the sufficient
condition ($b>1+\epsilon n_F$) for fixation of defection at node
1. Thus the (one time step) probability $Q_{DC}$ of invasion of a
cooperator node in $\cal F$ is

\begin{eqnarray}
Q_{DC} & = & \frac{1}{(n_F+1)} \frac{P_1-P_c}{\Delta (n_F+1)} \nonumber \\
 & & + \frac{(1-c)n_F}{(n_F+1)} \frac{P_d-P_c}{\Delta (n_F+1)}\;\;, \label{probCDter}
\end{eqnarray}
and the probability $Q_{CD}$ of invasion of a defector node in
$\cal F$ is
\begin{eqnarray}
Q_{CD} & = & \frac{1}{(n_F+1)} \frac{P_2-P_d}{\Delta
(n_F+n_C)}\;\;.
 \label{probDCter}
\end{eqnarray}

At time $t+1$, the expected fraction of cooperators is:

\begin{equation}
c(t+1) = c(t)(1-Q_{DC}) + (1-c(t))Q_{CD}\;\;.
\end{equation}

Assuming that the size of $\cal F$ is macroscopic, $n_F\gg1$, the
fraction of cooperators $c$ in $\cal F$ evolves according to the
differential equation

\begin{equation}
\dot{c} = (1-c)Q_{CD}-cQ_{DC}\;\;. \label{rate equation}
\end{equation}

Now, if $n_F\gg1$, and $n_C/(n_F)^2 \rightarrow 0$, then both
$Q_{CD}$ and the first term in the right-hand side of
(\ref{probCDter}) vanish, and we arrive to the differential
equation

\begin{equation}
\dot{c}= \frac{c(1-c)}{\Delta} (\epsilon(1-c)-(b-1)c)\;\;.
\label{diffeqbis}
\end{equation}

As expected, with a simple re-scaling of time, equation
(\ref{diffeqbis}) is no other than the replicator equation
(\ref{replicator}): note that in the limit $n_F\gg1$ that we have
considered, the probability that a node in $\cal F$ picks up a Big
Brother when updating its strategy is negligible, and then the
evolution inside the complete graph $\cal F$ is overwhelmingly
determined by the internal connections, and thus by the replicator
equation. In other words, in this limit of maximal possible
connectivity, BB1 and BB2 are no longer bigger than the nodes in
$\cal F$ and their influence on the fluctuating set is negligibly
small in the thermodynamical limit.

Note however that as far as $n_F$ is finite, the Theorem in
section \ref{dipole_net} still hold, and a chance for the
fluctuating activity inside $\cal F$ remains. We now turn
attention to situations where $k_F \ll n_F$, far from the social
panmixia.

\subsection{$\cal F$ is a disconnected graph (ideal-gas)}

\label{ideal gas}

Let us now obtain some explicit results for one of the simplest
choices for the topology of connections inside the fluctuating
set, namely $k_F=0$. In this case each node in $\cal F$ is only
connected to Big Brothers. This is in fact an effective single
node problem, where homogeneity assumption in $\cal F$ is exact;
in other words, the absence of internal interactions in the set
$\cal F$ is a sort of ideal-gas condition easy to deal with in the
large size limit.

\subsubsection{A differential equation for $c$}

Note that the sufficient conditions for fixation of defection at
BB1 and of cooperation at BB2 are respectively, $b > 1 +
\epsilon$, and $n_C> b-\epsilon n_F$. Denoting by $c(t)$ the
instantaneous fraction of cooperators in $\cal F$, the payoffs of
Big Brothers are given by (\ref{payoff12}), and the payoffs of a
cooperator node and a defector node in $\cal F$ are, respectively,
\begin{equation}
P_c= 1 + \epsilon \;\;\;\;\;\;\;,\;\;\;\;\;\;\;P_d=b\;\;.
\label{payoffCD}
\end{equation}

Then one finds for the (one time step) probability $Q_{DC}$ of
invasion of a cooperator node in $\cal F$
\begin{equation}
Q_{DC} = \frac{cb-(1+\epsilon)/n_F}{2\Delta}\;\;, \label{probCD}
\end{equation}
and using the notation $A=\epsilon + (n_C - b)/n_F$ and
$B=1+n_C/n_F$

\begin{equation}
Q_{CD} = \frac{A + c(1-\epsilon)}{2\Delta B}\;\;, \label{probDC}
\end{equation}
for the probability of invasion of a defector node in $\cal F$.
Note that $A>0$ due to the non-invasion of BB2 (sufficient)
condition.

Provided $n_F\gg1$, the fraction of cooperators $c$ in $\cal F$
evolves according to the differential equation (\ref{rate
equation}), which after insertion of expressions (\ref{probCD})
and (\ref{probDC}), and re-scaling of time, becomes

\begin{equation}
\dot{c} = f(c) \equiv A_0 + A_1 c + A_2 c^2\;\;, \label{diffeq}
\end{equation}
where the coefficients are

\begin{eqnarray}
A_0 & = & A\;\;, \\
A_1 & = & 1-\epsilon-A+B(1+\epsilon)/n_F\;\;, \\
A_2 & = & -(1-\epsilon+bB)\;\;, \label{quadraticcoeff}
\end{eqnarray}

One can easily check ($A_0>0$ and $A_2<0$) that there is always
one positive root $c^*$ of $f(c)$, which is the asymptotic value
for any initial condition $0\leq c(0) \leq 1$ of equation
(\ref{diffeq}). In this asymptotic regime, the one time step
invasion probabilities, $Q_{DC}$ and $Q_{CD}$, become time
independent and one can then compute the probability that the
cooperation strategy remains for a time $\tau_c\geq 1$ (permanence
time of cooperation) at a fluctuating node, simply as $P(\tau_c) =
Q_{DC}(1-Q_{DC})^{\tau_c-1}$. In a similar way, the distribution
density $P(\tau_d)$ of defection permanence times is obtained as
$P(\tau_d) = Q_{CD}(1-Q_{CD})^{\tau_d-1}$. Thus the distribution
densities of both strategies permanence time are exponentially
decreasing.

For $\epsilon=0$, in the so called weak PD game ({\em i.e.} at the
border between the PD and the Hawks and Doves game), if one
further assumes that the relative size $\mu(F)$ of the component F
is large enough, {\em i.e.} $\mu(F)\rightarrow 1$, and
$\mu(C)\rightarrow 0$, one easily obtains that the stationary
solution of equation (\ref{diffeq}) behaves as $c^* \simeq
(b+1)^{-1}$ near the limit $\mu(F)\rightarrow 1$. The distribution
density $P(\tau_c)$ of the cooperation permanence times of a
fluctuating node, as a function of the parameter $b$ is thus
\begin{equation}
P(\tau_c) = (2b+1)^{-1} \left(\frac{2b+1}{2b+2}\right)^{\tau_c},
\label{Ptau_c}
\end{equation}
and the distribution density $P(\tau_d)$ of defection permanence
times
\begin{equation}
P(\tau_d) = (2b(b+1)-1)^{-1}
\left(\frac{2b(b+1)-1}{2b(b+1)}\right)^{\tau_d} \label{Ptau_d},
\end{equation}

These distribution densities characterize the pace of invasion
cycles at a fluctuating node in the (asymptotic) equilibrium
strategic macroscopic state.

\subsubsection{A formal thermodynamical approach}

From the point of view of the set $\cal F$, when $n_F \gg 1$, the
model corresponds to a non-interacting (ideal) set of independent
phenotypic strategists that fluctuate due to a polar field (Big
Brothers influence) whose strength is self-consistently determined
by the average cooperation $c$. This problem is equivalent to that
of an ideal paramagnetic salt in a noisy (telegraphic) magnetic AC
field of intensity proportional to the average magnetization.

A typical and correct statistical-physicists approach "from
scratch" to this two-states model is the familiar micro-canonical
setting: At (dynamical) macroscopic equilibrium, the probability
of each strategic micro-state $l=\{s_i\}$ of fixed value of
$c_l=c$ is uniform
\begin{equation}
P_l = \Omega ^{-1}\;\;,
\end{equation}
where $\Omega=n_F!/((cn_F)!(n_F-cn_F)!) $) is their number. The
lack of information $S=\ln\Omega$ of the macro-state as a function
of global cooperation $n_Fc$, {\em i.e.} the relation $S(n_Fc)$,
can be regarded as the micro-canonical fundamental
"thermodynamical" relation, and its first derivative is the
intensive parameter $\beta$, that after using Stirling's
approximation is easily obtained as
\begin{equation}
\beta= \ln \left(\frac{1-c}{c}\right) \label{beta(c)}\;\;.
\end{equation}

This is an equation of state, which simply expresses the
connection of the equilibrium value of the macroscopic cooperation
level $c$ to the "entropic" intensive parameter $\beta$. Note that
$c$ is determined by the balance condition ($\dot{c}=0$):

\begin{equation}
\frac{1-c}{c}=\frac{Q_{DC}}{Q_{CD}}\;\;,
\end{equation}
from where equation of state (\ref{beta(c)}) determines $\beta$ as
a function of model parameters ({\em i.e.} $b$, $\epsilon$, and
$n_C/n_F$). For example, when $\epsilon =0$, $\beta = \ln b>0$,
indicating that the disorder of the activity increases with
increasing cooperation. The maximal value of $\beta \rightarrow
\infty$ corresponds to zero disorder ($b \rightarrow \infty$),
while its minimal zero value corresponds to highest possible value
(at $b=1$) of cooperation ($c=(1/2)$). Note that lower values of b
($b<1$) correspond to negative $\beta$ values, where entropy
decreases with increasing values of cooperation, outside the PD
domain (Stag Hunt game domain, see \cite{Roca}).

An alternative (and equivalent in the thermodynamic limit) setting
is to consider the whole space of $2^{n_F}$ configurations
$l=\{s_i \}_{i=1}^{n_F}$, of unrestricted $c_l$, but under the
condition that the average value $c=\sum_l {\cal P}_l c_l$ is
fixed. That is the familiar canonical setting. The normalization
factor $Z=\sum_l \exp(-\beta c_l)$ is the canonical partition
function (Boltzmann's Zustandsumme), that due to independence
($k=0$) is easily factorized as $(1+\exp(-\beta))^{n_F}$.

In the canonical setting a most informative macroscopic quantity
is the "heat capacity" analog: The fluctuations of $c_l$ along
representative (typical) stochastic trajectories at equilibrium
under the evolutionary dynamics of the game are, following the
standard thermodynamical formalism given by $\partial c/\partial
(\beta^{-1})$, so that this quantitative social indicator detects
very precisely sudden variations of macroscopic cooperation with
payoff's parameters. In this ideal-gas kind of case there are no
critical points and fluctuations do not diverge. For example, for
$\epsilon=0$ they are given by the (Bernouillian) binomial
variance $n_Fc(1-c)= n_Fb/(b+1)^2$.

\subsection{$\cal F$ is a random regular graph}
\label{mean-field}

Random regular networks are random networks of fixed degree $k$.
All nodes being thus equivalent, a sensible approach is to assume
(mean-field like) that the fraction of instantaneous cooperators
in the neighborhood of a node is the fraction $c$ of the whole set
$\cal F$. In other words, one neglects local fluctuations of $c$.
The contribution of the internal interactions to the variation of
$c$ is then of the "replicator equation" type, as discussed above
for the complete graph case. The difference here is that if $k_F
\ll n_F$ the contribution of the interactions with Big Brothers
cannot be longer neglected.

\subsubsection{Mean-field approximation}

The payoffs of Big Brothers BB1 and BB2 are given by equation
(\ref{payoff12}), and the payoffs of a cooperator node and a
defector node at $\cal F$ under the mean-field assumption are:
\begin{equation}
P_c= ck+1+\epsilon(k(1-c)+1) \;\;\;,\;\;\;\;\;P_d=(ck+1)b\;\;.
\label{payoffCDbis}
\end{equation}

\begin{figure}
\begin{center}
\epsfig{file=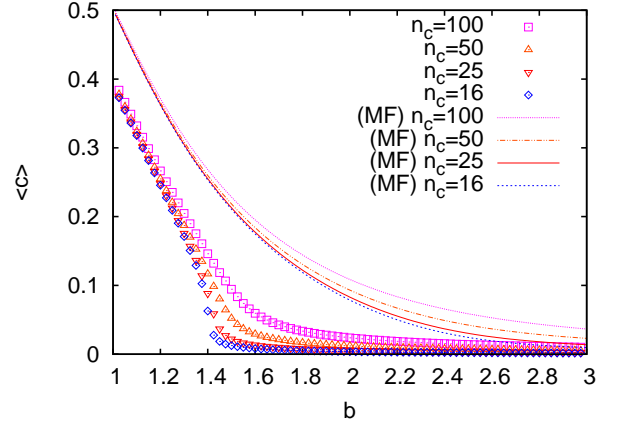,width=2.3in,angle=-90,clip=1}
\end{center}
  \caption{Macroscopic cooperation in a random regular graph structure
  for the set ${\cal F}$, with $k=4$, and $n_F=4000$, and $\epsilon=0$.
  A decreasing sequence of $n_C/n_F$, as indicated in figure, has been
  used. Symbols represent numerical Monte Carlo results, and the different
  lines represent the mean-field predictions as given by the solution ($\dot{c}=0$) of Eq.\ (\ref{diffeq2}).}
\label{randonet}
\end{figure}

The differential equation for $c$ is then
\begin{eqnarray}
\dot{c} & = & \frac{(1-c)(P_2-P_d)}{(k+2)Bn_F\Delta}- \frac{c(P_1-P_c)}{(k+2)n_F\Delta} \nonumber \\
 & & + \frac{(1-c)ck(P_c-P_d)}{(k+2)^2\Delta}\;\;, \label{rreq}
\end{eqnarray}
which under the assumption $kb \ll n_F$, takes the form
\begin{equation}
\dot{c} = f(c) \equiv \frac{1}{(k+2)^2B\Delta}(A'_0 + A'_1 c +
A'_2 c^2 + A'_3 c^3)\;\;, \label{diffeq2}
\end{equation}
where the coefficients are

\begin{eqnarray}
A'_0 & = & (k+2)(B-1+\epsilon)\;\;, \\
A'_1 & = & 2(2(1-\epsilon)-B)+k(2(1-\epsilon)-B(b-\epsilon)) \nonumber \\
     &   & +k^2B\epsilon\;\;, \\
A'_2 & = & 2(\epsilon-1-Bb)+k(\epsilon-1-B(1+\epsilon)) \nonumber \\
     &   & +k^2B(1-b-2\epsilon)\;\;, \\
A'_3 & = & k^2B(b-1+\epsilon)\;\;, \label{quadraticcoeff}
\end{eqnarray}

Note that the assumption $n_C > b-n_F\epsilon$ ({\em i.e.} the
condition for Big Brother 2 to be a permanent cooperator) implies
that $A'_0
> 0$, so that $\dot{c}(0) >0$ and one positive root, say $c^*$, of $f(c)$ is
then ensured, in agreement with the theorem of section
\ref{dipole_net}.

\subsubsection{A social phase transition, and the mean-field
failure.}

\label{binomial}

We now adopt the statistical mechanics formal perspective, and
proceed to explicitly compute the equilibrium macro-state, {\em
i.e.} the stationary probability distribution density ${\cal
P}^*_l $, inside the mean-field approximation.

Let us consider two different (arbitrary) strategic microstates
$l= \{s_i\}$ ($i=1, ..., n_F$), and $l'=\{s'_i\}$, of the
fluctuating set. For any pair of microstates $(l,l')$ we define
the following numbers:

\begin{eqnarray}
n_{11} & = & \sum_i \delta_{s_i,s'_i} \delta_{s'_i,1}\;\;,\\
n_{10} & = & \sum_i (1-\delta_{s_i,s'_i}) \delta_{s'_i,0}\;\;,\\
n_{00} & = & \sum_i \delta_{s_i,s'_i} \delta_{s'_i,0}\;\;,\\
n_{01} & = & \sum_i (1-\delta_{s_i,s'_i}) \delta_{s'_i,1}\;\;,
\label{n's}
\end{eqnarray}
{\em i.e.}, $n_{11}$ is the number of nodes that are cooperators
in both microstates, $n_{10}$ that of the nodes that are
cooperators in $l$ but defectors in $l'$, etc... Using equation
(\ref{cooperation}) it is straightforward to obtain
\begin{equation}
c_l - c_{l'}= \frac{1}{n_F} (n_{10} - n_{01})\;\;.
\label{canonical1}
\end{equation}

Now, let us assume that the probabilities that a node $i$ changes
strategy are independent of node $i$ (homogeneity assumption,
mean-field), and denote them by $Q_{CD}$ (transition from defector
to cooperator) and $Q_{DC}$ (for the transition from cooperator to
defector). Then we can easily see that the transition
probabilities between the microstates $l$ and $l'$ are given by

\begin{eqnarray}
\Pi_{l,l'} & = & (1-Q_{DC})^{n_{11}} (1-Q_{CD})^{n_{00}} Q_{DC}^{n_{01}} Q_{CD}^{n_{10}}, \\
\Pi_{l',l} & = & (1-Q_{DC})^{n_{11}} (1-Q_{CD})^{n_{00}}
Q_{DC}^{n_{10}} Q_{CD}^{n_{01}},
\end{eqnarray}

Henceforth, denoting $\exp(-\beta)= Q_{CD}/Q_{DC}$, one easily
obtains the expression:

\begin{equation}
\Pi_{l,l'} \exp(-\beta c_{l'}n_F) = \Pi_{l',l} \exp(-\beta c_l
n_F)\;\;, \label{detailed balance}
\end{equation}
from where the unique solution to the fixed point equation

\begin{equation}
\Pi_{l,l'} {\cal P}^*_{l'} = {\cal P}^*_l\;\;, \label{fixed point}
\end{equation}
is easily found to be:
\begin{equation}
{\cal P}^*_{l} = Z^{-1} \exp(-\beta c_l n_F)\;\;, \label{canonical
measure}
\end{equation}
where $Z$ is the canonical partition function
\begin{equation}
Z = \left[ \frac{Q_{CD}+Q_{DC}}{Q_{DC}} \right]^{n_F}\;\;.
\label{partition function}
\end{equation}

As it is well-known {\cite{Jaynes}}, the canonical probability
distribution density (\ref{canonical measure}) is the unique
density that maximizes the lack of information (entropy),
$S=-\sum_l{\cal P}_l \ln {\cal P}_l$, among those (compatible)
densities that share a common value for the macroscopic average of
cooperation $c=\sum_l {\cal P}_l c_l$. This provides a standard
thermodynamic meaning to the parameter $\beta$: it is no other
than the intensive entropic parameter associated to cooperation,
that is, the Lagrange multiplier \cite{Courant-Hilbert,Tisza}
associated to the restriction $c=\sum_l {\cal P}_l c_l$ on the
compatible measures (canonical restricted maximization of
entropy), that is:

\begin{equation}
\beta = \frac{\partial S}{n_F\partial c}\;\;. \label{social
temperature}
\end{equation}

The parameter $\beta$ simply measures how fast the entropy of the
equilibrium macro-state increases versus global cooperation
variations. Its formal role is that of inverse thermodynamical
temperature. The fluctuations of the micro-states cooperation
$c_l$, namely $(n_F)^2 (\sum_l({\cal P}_lc_l^2) - (\sum_l{\cal
P}_lc_l)^2)$ are given by $ n_Fc(1-c)$. The dependence on the game
and network parameters $b, \epsilon, n_C/n_F, k$ of the
fluctuations of cooperation is obtained by solving for the
cooperation equilibrium value $\dot{c}=0$ in (\ref{diffeq2}), and
plotted in figure for $k=4$, $\epsilon=0$, and decreasing values
of the ratio $n_C/n_F$.

\begin{figure}
\begin{center}
\epsfig{file=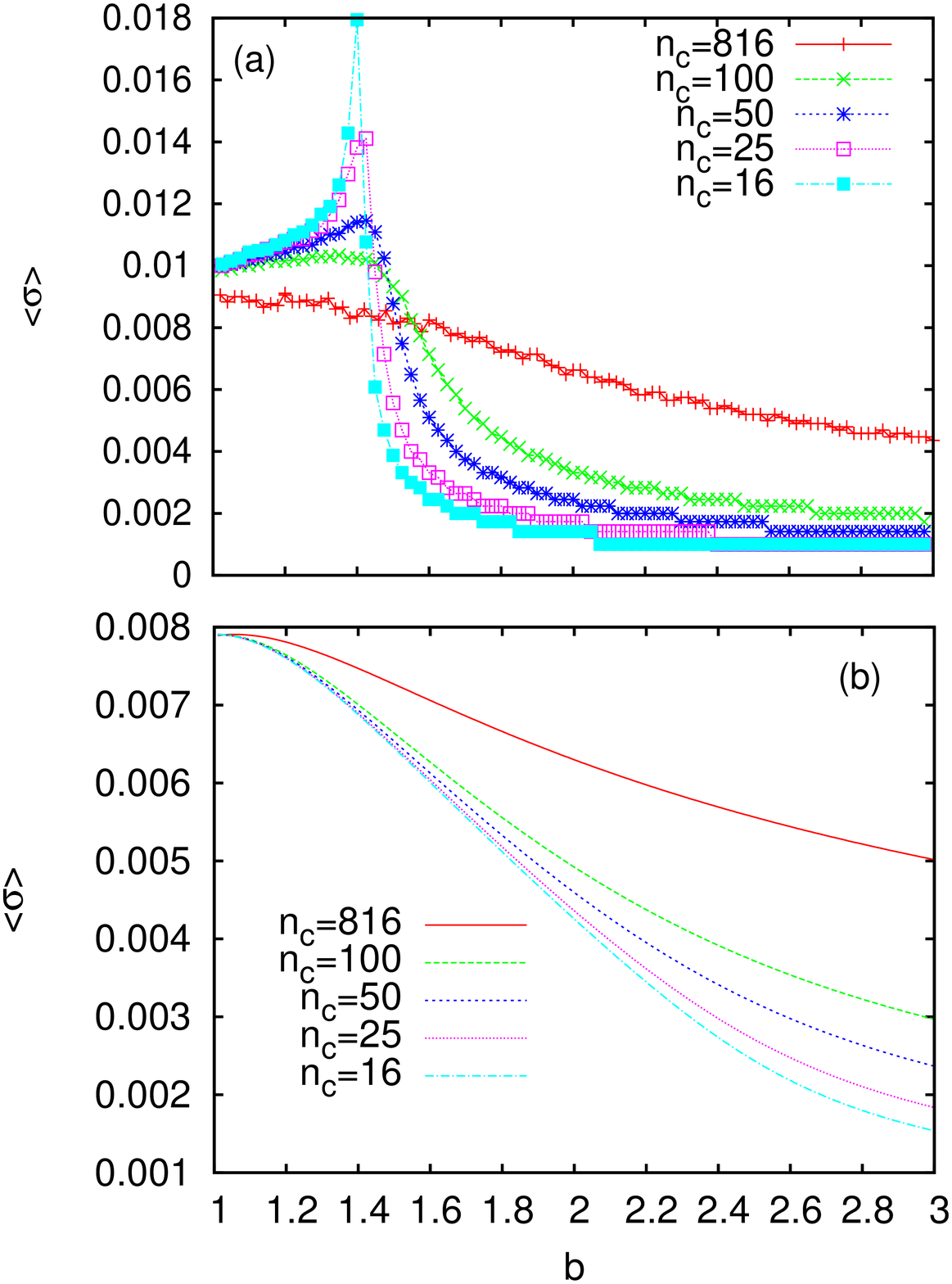,width=3.3in,angle=-0,clip=1}
\end{center}
  \caption{Fluctuations of cooperation in a random regular graph
  structure for the set ${\cal F}$. The upper panel (a) shows, for $k=4$,
  $\epsilon=0$, $n_F=4000$, and a decreasing sequence of $n_C/n_F$ values as indicated,
  the fluctuations of cooperation observed in Monte Carlo simulations. The
  lower panel (b) shows the mean-field predictions. The mean-field approach
  is shown in text to be unable to predict the observed phase transition. This
  qualifies network reciprocity as a true "critical" social phenomenon.}
\label{randonet}
\end{figure}

To which extent the mean-field prediction fails for low values of
the parameter $n_C/n_F$, can be appreciated by confronting the
prediction above with the results from Monte Carlo simulations.
There in figure we see how a peak in cooperation fluctuations is
revealed, when $n_C/n_F \rightarrow 0$, signaling the occurrence
of a phase transition between two qualitatively different
macroscopic behaviors, that correspond to low and high temptation
regimes. The mean-field assumption is thus only valid if the
payoff received from ${\cal C}$ by Big Brother 2 is not negligible
versus the size $n_F$.

The reasons for the failure of the mean-field approximation rely
on the lattice reciprocity of internal interactions, which is
totally absent in the mean field approximation. Let us remind here
our remark above on the replicator-equation-type of effect of
internal interactions in equation (\ref{diffeq}) because of the
mean-field assumption. The transition signaled by the divergence
of fluctuations at $b^*$ reveals the onset of internal lattice
reciprocity, a conclusion that we now substantiate (see also
appendix \ref{appendix} below).

For $b>b^*$, say in the low-temperature (high temptation) phase,
the macro-state is dominated by fast defection invasions on the
relatively few nodes that are instantaneous cooperators due to
sporadic interactions with Big Brother 2. In the appendix
\ref{appendix} we show that, in the low $c$ and low $n_C/n_F$
regime, the BB-imitation events in a given node are typically
separated by intervals of time of about $c^{-1}$ time units large.
In those large intervals when Big Brother 2's influence is null,
the very few and mostly isolated instantaneous cooperators are
quickly invaded by defector internal neighbors. In this regime
lattice reciprocity has no chance to develop, and cooperation is
only weakly sustained by the sporadic influence of BB2.

On the contrary, for $b<b^*$ (high temperature, or low temptation
phase) the local fluctuations of the neighbors strategic field
favor the building up of clusters of cooperators that resist
invasions during time intervals that are comparable to the
characteristic time intervals between BB-imitation events. Under
these circumstances the "extra payoff" that BB2 receives from
${\cal C}$ does not anymore need to be high in order to sustain
high levels of cooperation. Internal lattice reciprocity enhances
the probability of highly cooperative micro-states, so that the
macro-states below transition differs substantially from those of
the high-temptation phase. This was not captured by the mean-field
approximation, for these effects require a sizable likelihood of
occurrence for the local fluctuations of the strategic field, and
the neglect of them is all a mean-field approach is based upon.

To summarize the discussion of the results shown in figure, a
random regular structure of interactions inside ${\cal F}$ is
enough to support lattice reciprocity mechanisms that cannot be
captured by a simple mean-field approach. The onset of lattice
reciprocity is furthermore interpreted as a "thermodynamical"
phase transition, in a rigorous formal sense (divergence of the
fluctuations of an extensive parameter, the cooperation $c$). One
is then lead to a sensible and precise formal framework where such
a term as "social temperature" is not a vague metaphor, but it
denotes a truly quantitative parameter, a legitimate (measurable,
observable) social indicator.

\section{Prospective remarks}
\label{IV}

The plausibility of a thermodynamical perspective on Evolutionary
game dynamics studies is not a new issue, for it is somehow
implicit (or at least connatural) to a body of research literature
on statistical mechanics of strategic interactions
\cite{szabo,blume}. What our simple analysis here shows is that it
can sometimes be strengthened up to a formal interpretation of
quantitative macroscopic social indicators as thermodynamic
quantities. In the extent that it helps to understand and to
quantitatively characterize the phenomenology of social and
economical models, it should be recognized as a powerful
theoretical perspective. What is even more important, this
perspective emphasizes the central role of quantitative
(experimental, observational) studies in social sciences, and
could provide, in those contexts, alternate valuable meanings to
quantitative social indicators and even suggestions for new and
better ones.

Any "general-physics" trained scientist recognizes that entropy
reasoning is an extraordinary powerful tool for the analysis of
macroscopic behavior in (material) traditional-physics systems. It
turns out that some of the models (at least a bunch of interesting
ones) of social phenomena are to a large extent amenable to a
macroscopic description where thermodynamical concepts have proved
to be essential. Of course, some notions like {\em e.g.} "First
Law of Thermodynamics" could be often absent in these new
contexts. However we emphasize that the absence of energy as a
variable in social models is not a shortcoming for the
applicability {\em mutatis mutandi} of many aspects of the
thermodynamical formalism to these models. A word of caution is
nevertheless worth here regarding typical system sizes in
controlled social experiments, where finite size effects could be
hugely determinant.

Nowadays, it is somewhat generally accepted that Physics in
general, and Statistical Physics in particular, offers a powerful
tool-box for problem solving in Social Sciences and many other
areas. Recent trends in cognitive science \cite{Page} have
correctly emphasized the power of the "diversity of perspectives"
in problem solving, so it does not come as a surprise that adding
physical perspectives to social models may sometimes pave the way
to the needed breakthrough. Perhaps one should also wonder about
the possibility of reverse flow in these interdisciplinary
approaches to Social Sciences. After all, the proper use of a tool
helps to its reshaping, and one could perhaps expect some kind of
feedback. In other words, is there any new physics that we can
learn from the study of Social and Economic Complex Systems? Only
the recourse to empirical and quantitative methods in the study of
social phenomena may likely give clues for sensible answers to
this question.

\begin{acknowledgments}
  LMF acknowledges the organizers and attendants to the ISI
  workshop on Socio-Physics (Torino, Italy, May 2008) for the
  inspiring scientific atmosphere they created, and the warm
  (heated, in some case) reception given to the main contents
  of this paper. Discussions with B. Roehmer helped us to shape
  the final section on prospects. J. G.-G. and Y. M. are supported by MCyT through the Juan de la
  Cierva and the Ram\'{o}n y Cajal Programs, respectively. This work has been partially supported by
  the Spanish DGICYT Projects FIS2006-12781-C02-01, and FIS2005-00337, and a DGA grant to FENOL group.
\end{acknowledgments}

\begin{appendix}

\subsection{ Low $c$ approximation}
\label{appendix}

In order to simplify expressions we assume hereafter $\epsilon=0$
and $k=4$, and denote $\delta=n_C/n_F$. For the case of a random
regular graph structure of the fluctuating set ${\cal F}$, the
probability $\Pi_{C\leftarrow D}^{BB}$ that a instantaneous
defector node chooses to imitate Big Brother 2 (invasion event
from BB2) is, to first order in $n_F^{-1}$,

\begin{equation}
\Pi_{C\leftarrow D}^{BB}= \frac{1}{(k+2)} \frac{c+\delta}
{(1+\delta)b} \label{A1}
\end{equation}
while the probability $\Pi_{D\leftarrow C}^{BB}$ of an invasion
event from BB1 to an instantaneous cooperator node in ${\cal F}$
is, to first order in $n_F^{-1}$,

\begin{equation}
\Pi_{D\leftarrow C}^{BB}= \frac{c}{(k+2)} \label{A1}
\end{equation}

Thus, for $\delta \leq c$, typical intervals between invasion
events from Big Brothers in a node are (of the order of) $c^{-1}$
time units large. For large values of the temptation, where the
value of $c$ is expected to be very small, the dynamics is
consequently dominated, for typically very large intervals of
time, by internal strategic interactions. Let us analyze them.

The internal neighbors of a cooperator $i$ are overwhelmingly
likely instantaneous defectors in this "low $c$" regime, so that
$i$ will be quickly invaded by them. The only chance for it to
resist invasion would be that its instantaneous neighborhood
microstate had at least two cooperator neighbors and that
$b<(3/2)$ (note that in this strategic configuration, the payoff
of $i$ is $P_i=3$ and that of its typical defector neighbors is
$2b$). These neighborhood microstates (cooperative clusters) are
so rare fluctuations that low values of the temptation $b$ are
necessary for their non-negligible occurrence. Provided $b$ is
below the transition value, the resilience to invasion (lattice
reciprocity) of cooperative clusters enhances the likelihood of
these fluctuations, which in turn reinforces the clusters
resilience, and so on. This positive feedback mechanism of
cooperative fluctuations enhancement is thus what triggers the
transition to highly cooperative macro-states, and qualifies
lattice reciprocity as a critical social phenomenon.

\end{appendix}

\end{document}